\documentclass[11pt]{article}
\usepackage[fleqn]{amsmath}

\usepackage{amssymb}
\usepackage{graphicx}

\newtheorem{theorem}{Theorem}[section]
\newtheorem{corollary}{Corollary}[section]
\newtheorem{conj}{Conjecture}[section]

\def\x{$\hfill\rlap{$\sqcup$}\sqcap$\bigskip}

\def\atn#1{\setbox10=\vbox{#1}\par
\hbox{\vbox{\hbox to 0pt{\hss\vrule width 1pt height \ht10
\ \vrule width 1pt height \ht10\kern 5pt}}\box10}}

\def\f#1{\mathbb{F}_{#1}}
\def\no{n$^\circ$}

\begin{document}

\title{Some More  Functions\\ That Are Not APN Infinitely Often.\\ The Case of Kasami exponents}
\author{Fran\c cois Rodier
}
\date{}
\maketitle

\begin{abstract}
We prove a necessary condition for some polynomials of Kasami degree  to be   APN over $\mathbb{F}_{q^n}$ for large $n$.
\end{abstract}

\section{Introduction}

The vector Boolean functions are used in cryptography to
construct block ciphers and an important criterion on these
functions is high resistance to differential cryptanalysis.

Let  $q=2^n$ for some positive integer $n$.
A function $f~:~\f q \longrightarrow \f q$ is said to be \emph{almost perfect nonlinear} (APN) on $\f q$
if the number of solutions in $\f q$ of the equation
$$f(x+a)+f(x)=b$$
is at most 2, for all $a,b\in \f q$, $a \not=0$. 
Because $\f q$ has characteristic 2, the number of solutions
to the above equation must be an even number, for any function $f$ on $\f q$.
This kind of function has a good resistance to differential cryptanalysis as was  proved by  Nyberg in \cite{ny}.

So far, the study of APN functions has focused on power functions.
Recently it was generalized to polynomials (cf. \cite{amr}).

There are many classes of function for which it can be shown
that each function is APN for at most a finite number of extensions. 
 So we fixe a finite field $\f{q}$ and a function $f:\f{q}\to \f{q}$ given by a polynomial in $\f{q}[x]$ and
we set the question of whether this function can be APN  for an infinite number of extensions of $\f{q}$.
 
In this approach, Hernando and McGuire  \cite{HM} showed a result on the classification of APN monomials which has been conjectured for 40 years: the only exponents such that the monomial $x^d$ are APN over  infinitely many 
extension of $\f2$ are of the form $2^i+1$ or $4^i-2^i+1$. One calls these exponents {\sl exceptional exponents}.
Then it is natural to 
 formulate for polynomial functions
the following conjecture.
\begin{conj}[Aubry, McGuire and Rodier]
A polynomial on $\f q$\break can
be APN for an infinity of extensions of $ \f q $ only if it is CCZ equivalent (as was defined by Carlet, Charpin and   Zinoviev in \cite{ccz}) to a monomial $ x ^ t $
where $ t $ is an exceptional exponent. 
\end{conj}

Some cases for $f$ of small degree have been proved by the author \cite{FR}.
We showed there that for some polynomial functions $f$  which are APN on $\f2^m$, the number $ m $ is bounded by an expression depending on the degree of $ f $.

We used it for a method already used by Janwa who showed, with the help of Weil bounds,  that certain cyclic codes could not correct two errors \cite {JW}.
Canteaut showed by the same method that some power functions were not APN for a too large value of the exponent \cite {AC}.
We were able to generalize this result to all polynomials by applying  Lang-Weil's results.

Some cases of this conjecture have been studied already, in particular the case of Gold degree. We recall them in section \ref{rappel}.
In this paper, we will study polynomials of Kasami degree.
The proofs happen to be somehow the same as in Gold degree, with a few changes anyway.

\section{Preliminaries}

We define
\[
\phi (x,y,z) = \frac{f(x)+f(y)+f(z)+f(x+y+z)}{(x+y)(x+z)(y+z)}
\]
which is a polynomial in $\mathbb{F}_q [x,y,z]$.  This  polynomial defines a surface $X$ in the three dimensional affine space $ \mathbb{A}^3$.

If $X$ is absolutely irreducible (or has an absolutely irreducible component
defined over $\mathbb{F}_q$) then $f$ is not APN on 
$\mathbb{F}_{q^n}$ for all $n$ sufficiently large.
As shown in \cite{FR}, this follows from the Lang-Weil bound for surfaces, which guarantees many
$\mathbb{F}_{q^n}$-rational points on the surface for all $n$ sufficiently large.

We call $\phi_j (x,y,z)$ the $\phi$ function associated to the monomial $x^j$. The function $\phi_j (x,y,z)$ is is homogeneous of degree $j-3$.

We recall a result due to  Janwa, Wilson, \cite[Theorem 5]{JW} about Kasami exponents.
 
 \begin{theorem}
 \label{Kasami}
If $f(x)=x^{2^{2k}-2^k+1}$ then
 \begin{equation}\label{Goldfactors}
 \phi(x,y,z)=\prod_{\alpha\in \mathbb{F}_{2^k}-\mathbb{F}_2}p_\alpha(x,y,z)
 \end{equation}
 where for each $\alpha$, $p_\alpha(x,y,z)$ is an absolutely irreducible polynomial of degree $2^k+1$ on $\mathbb{F}_{2^k}$ such that
 $p_\alpha(x,0,1)=(x-\alpha)^{2^k+1}$.
\end{theorem}

\section{Some  Functions That Are Not APN Infinitely Often}
\label{rappel}

The best known examples of APN functions are the Gold functions $x^{2^k+1}$ and
the Kasami-Welch functions $x^{4^k-2^k+1}$.
These functions are defined over $\mathbb{F}_{2}$, and are APN on 
any field $\mathbb{F}_{2^m}$ where $gcd(k,m)=1$.
For other odd degree polynomial functions, we can state a general result.

\begin{theorem}[Aubry, McGuire and Rodier, \cite{amr}]
If the degree of the polynomial function $f$ is odd and not a Gold or a Kasami-Welch number then $f$ is not APN over $\mathbb{F}_{q^n}$ for all $n$ sufficiently large. 
\end{theorem}

In the even degree case, we can state the result when half of the degree is odd, with an extra minor condition.

\begin{theorem}[Aubry, McGuire and Rodier, \cite{amr}]
\label{degre2e}
If the degree of the polynomial  function $f$ is $2e$
with $e$ odd,   and if $f$ contains a term of odd degree, then $f$ is not APN over $\mathbb{F}_{q^n}$ for all $n$ sufficiently large. 
\end{theorem}

In \cite{FR1} we have some results for the case of polynomials of degree $4e$ where $e$ is odd.

\begin{theorem}
\label{geom}
If the degree of the polynomial function $f$ is even such that
$\deg(f)=4e$ with $e\equiv3\pmod4$,  
and if the polynomials of the form 
$$(x+y)(y+z)(z+x)+P$$ with 
\begin{equation}
\label{poly}
P(x,y,z)= c_1 (x^2 + y^2 + z^2) + c_4 (x y+   x z+  z y)+ b_1 (x+  y+   z)+d
\end{equation}
for $c_1, c_4, b_1, d\in \f{q^3}$,
do not divide $\phi$
then $f$ is not APN over $\mathbb{F}_{q^n}$ for $n$ large. 
\end{theorem}

We have more precise results for polynomials of degree 12.
\begin{theorem}
\label{d12}
If the degree of the polynomial $f$ defined over $\f q$ is 12,    then either $f$ is not APN over $\mathbb{F}_{q^n}$ for large $n$ or $f$ is CCZ equivalent to the Gold function $x^3$. 
{In this case $f$ is of the form 
$$L(x^3)+L_1 \hbox{ or } (L(x))^3+L_1$$
where $L$ is a linearized polynomial 
$$x^4+x^2(c^{1+q}+c^{1+q^2}+c^{q+q^2})+x c^{1+q+q^2},$$
$c$ is an element of $\f{q^3}$ such that $c+c^q+c^{q^2}=0$ and $L_1$ is a q-affine polynomial of degree at most 8 (that is a polynomial   whose monomials are of degree  0 or a power of 2).}
\end{theorem}

We have some results on the polynomials of Gold degree $d=2^k+1$.

\begin{theorem}[Aubry, McGuire and Rodier, \cite{amr}]
Suppose $f(x)=x^d+g(x)$ where $\deg (g) \leq 2^{k-1}+1$ .
Let $g(x)=\sum_{j=0}^{2^{k-1}+1} a_j x^j$.
Suppose moreover that there exists a nonzero coefficient $a_j$ of $g$ such that 
$\phi_{j} (x,y,z)$ is absolutely irreducible (where $\phi_i (x,y,z)$ denote the polynomial $\phi (x,y,z)$ associated to $x^i$).
Then   $f$ is not APN over $\mathbb{F}_{q^n}$ for all $n$ sufficiently large. 
\end{theorem}

\section{Polynomials of Kasami Degree}

Suppose the degree of $f$ is a Kasami number $d=2^{2k}-2^k+1$. 
Set $d$ to be this value for this section.
Then the degree of $\phi$ is $d-3=2^{2k}-2^k-2$.
We will prove the absolute irreducibility for a certain type of $f$.

\begin{theorem}
Suppose $f(x)=x^d+g(x)$ where $\deg (g) \leq 2^{2k-1}-2^{k-1}+1$ .
Let $g(x)=\sum_{j=0}^{2^{k-1}+1} a_j x^j$.
Suppose moreover that there exists a nonzero coefficient $a_j$ of $g$ such that 
$\phi_{j} (x,y,z)$ is absolutely irreducible.
Then $\phi (x,y,z)$ is absolutely irreducible.
\end{theorem}

Proof:  
Suppose $\phi(x,y,z)=P(x,y,z)Q(x,y,z)$ with $\deg P\ge \deg Q$.
Write each polynomial as a sum of homogeneous parts:
\begin{equation}\label{homogeneous}
\sum_{j=3}^{d} a_{j}\phi_j (x,y,z)= (P_s+P_{s-1}+\cdots +P_0)(Q_t+Q_{t-1}+\cdots +Q_0)
\end{equation}
where $P_j, Q_j$ are homogeneous of degree $j$.
Then from the Theorem (\ref{Kasami}) we get
\[
P_sQ_t=\prod_{\alpha\in \mathbb{F}_{2^k}-\mathbb{F}_2}p_\alpha(x,y,z).
\]
In particular this implies that  $P_s$ and $Q_t$ are relatively prime as the product is made of distinct irreducible factors.

The homogeneous terms of degree less than $d-3$ and greater than $2^{2k-1}-2^{k-1}$ are 0,
by the assumed bound on the degree of $g$.
Equating terms of degree $s+t-1$ in the equation (\ref{homogeneous}) gives
$P_s Q_{t-1}+P_{s-1}Q_t=0$.  Hence $P_s$ divides $P_{s-1}Q_t$
which implies $P_s$ divides $P_{s-1}$
because $gcd(P_s,Q_t)=1$, and we conclude $P_{s-1}=0$ as $\deg P_{s-1}<\deg P_{s}$.
Then we also get $Q_{t-1}=0$.
Similarly, $P_{s-2}=0=Q_{t-2}$, $P_{s-3}=0=Q_{t-3}$, and so on until we get the equation
\[
P_s Q_0+P_{s-t}Q_t=0
\]
since we suppose that $s\geq t$.
This equation implies $P_s$ divides $P_{s-t}Q_t$, which implies $P_s$ divides
$P_{s-t}$, which implies $P_{s-t}=0$.
Since $P_s\not=0$ we must have $Q_0=0$.

We now have shown  that $Q=Q_t$ is  homogeneous.
In particular, this means that $\phi_j (x,y,z)$ is divisible by $p_\alpha(x,y,z)$ for
some $\alpha \in  \mathbb{F}_{2^k}- \mathbb{F}_2$ and for all $j$ such that $a_{j}\ne0$.
We are done if there exists such a $j$ with $\phi_j (x,y,z)$ irreducible. Since $\phi_j (x,y,z)$ is defined over $\f2$ it implies that $p_\alpha(x,y,z)$ also, which is a contradiction with the fact that $\alpha$ is not in $\f2$.

\x

\noindent
{\bf Remark}:
The hypothesis that there should exist a $j$ with $\phi_{j} (x,y,z)$ is absolutely irreducible
is not a strong hypothesis.  This is true in many cases (see  remarks in \cite{amr}).
However, some hypothesis is needed, because the theorem is false without it.
One counterexample is with $g(x)=x^{13}$ and $k\geq 4$ and even.

\begin{corollary}
Suppose $f(x)=x^d+g(x)$ where $g$ is a polynomial in $ \mathbb{F}_q[x]$ such that $\deg (g) \leq 2^{2k-1}-2^{k-1}+1$ .
Let $g(x)=\sum_{j=0}^{2^{k-1}+1} a_j x^j$.
Suppose moreover that there exists a nonzero coefficient $a_j$ of $g$ such that 
$\phi_{j} (x,y,z)$ is absolutely irreducible.
Then the polynomial $f$ is APN for only finitely many extensions of $ \mathbb{F}_q$.

\end{corollary}

\subsection{On the Boundary of the First Case}

If we jump one degree more we need other arguments to prove irreducibility.

\begin{theorem}
Let $q=2^n$.
Suppose $f(x)=x^d+g(x)$ where $g(x)\in \mathbb{F}_q [x]$ and $\deg (g) = 2^{2k-1}-2^{k-1}+2$.
Let $k\ge3$ be odd and relatively prime to $n$. 
{If $g(x)$ does not have the form $ax^{2^{2k-1}-2^{k-1}+2}+a^2x^3$ then
$\phi$ is absolutely irreducible, while if
$g(x)$ does have the form $ax^{2^{2k-1}-2^{k-1}+2}+a^2x^3$
then either $\phi$ is irreducible or $\phi$ splits into two absolutely irreducible factors which are both
defined over $\mathbb{F}_q$.}
\end{theorem}

Proof:  
Suppose $\phi(x,y,z)=P(x,y,z)Q(x,y,z)$  with $\deg P\ge \deg Q$ and
let 
$$g(x)=\sum_{j=0}^{2^{2k-1}-2^{k-1}+2} a_j x^j.$$
Write each polynomial as a sum of homogeneous parts:
\[
\sum_{j=3}^{d} a_{j}\phi_j (x,y,z)= (P_s+P_{s-1}+\cdots +P_0)(Q_t+Q_{t-1}+\cdots +Q_0).
\]
Then
\[
P_sQ_t=\prod_{\alpha\in \mathbb{F}_{2^k}-\mathbb{F}_2}p_\alpha(x,y,z).
\]
In particular this means $P_s$ and $Q_t$ are relatively prime as in the previous theorem.

Since $s\geq t$, we have $s\geq 2^{2k-1}-2^{k-1}-1$.
Comparing each degree gives
$P_{s-1}=0=Q_{t-1}$, $P_{s-2}=0=Q_{t-2}$, and so on until we get the equation
of degree $s+1$
\[
P_{s} Q_1+P_{s-t+1}Q_{t}=0
\]
which implies $P_{s-t+1}=0=Q_1$.

If $s\not=t$ then $s\geq 2^{2k-1}-2^{k-1}$.
Note then that $a_{s+3} \phi_{s+3}=0$.
The equation of degree $s$ is
\[
P_{s} Q_0+P_{s-t}Q_{t}=a_{s+3} \phi_{s+3}=0.
\]
This means that $P_{s-t}=0$, so $Q_0=0$.
We now have shown  that $Q=Q_t$ is  homogeneous.
In particular, this means that $\phi(x,y,z)$ is divisible by $p_\alpha(x,y,z)$ for
some $\alpha \in  \mathbb{F}_{2^k}- \mathbb{F}_2$, which is impossible, as we will show.
Indeed, since the leading coefficient of $g$ is not 0, the polynomial $\phi_{2^{2k-1}-2^{k-1}+2}$ occurs in $\phi$;
as 
\begin{equation}
\label{decn}
\phi_{2^{2k-1}-2^{k-1}+2}= \phi^2_{2^{2k-2}-2^{k-2}+1}(x+y)(y+z)(z+x),
\end{equation}
 this polynomial is prime to $\phi$, because if $p_\alpha(x,y,z)$ occurs  in the polynomials $\phi_{2^{2k-1}-2^{k-1}+2}$, then it will occur in $\phi_{2^{2k-2}-2^{k-2}+1}$. 
If that is the case, the polynomial  $p_\alpha(x,0,1)=(x-\alpha)^{2^k+1}$ would divide $\phi_{2^{2k-2}-2^{k-2}+1}(x,0,1)$.
One has
$$\displaylines{(x+y)(y+z)(z+x)\phi_{2^{2k-2}-2^{k-2}+1}(x,y,z)\hfill\cr\hfill=x^{2^{2k-2}-2^{k-2}+1}+y^{2^{2k-2}-2^{k-2}+1}+z^{2^{2k-2}-2^{k-2}+1}+(x+y+z)^{2^{2k-2}-2^{k-2}+1}}$$
hence
$$x(x+1)\phi_{2^{2k-2}-2^{k-2}+1}(x,0,1)
=x^{2^{2k-2}-2^{k-2}+1}+1+(x+1)^{2^{2k-2}-2^{k-2}+1} .$$
Let $s=x-\alpha$.
We have, for some polynomial $R$:
\begin{eqnarray*}
&&(s+\alpha)(s+\alpha+1)s^{2^k+1}\\
&=&(s+\alpha)^{2^{2k-2}-2^{k-2}+1}+1+(s+\alpha+1)^{2^{2k-2}-2^{k-2}+1} \\
&=& \alpha^{2^{2k-2}-2^{k-2}+1}+s \alpha^{2^{2k-2}-2^{k-2}}+s^{2^{k-2}} \alpha^{2^{2k-2}-2^{k-1}+1}+1+\\
&&\qquad+(\alpha+1)^{2^{2k-2}-2^{k-2}+1}+s(\alpha+1)^{2^{2k-2}-2^{k-2}}+s^{2^{k-2}}(\alpha+1)^{2^{2k-2}-2^{k-1}+1} +s^{2^{k-2}+1}R(s).
\end{eqnarray*}
As
$\alpha^{2^k-1}=1$ we have
$ \alpha^{2^{2k-2}-2^{k-2}}= \alpha^{2^{k-2}(2^{k}-1)}=1$.
So
\begin{eqnarray*}
&&(s+\alpha)(s+\alpha+1)s^{2^k+1}\\
&=&\alpha+s+s^{2^{k-2}}\alpha^{1-2^{k-2}}+1+(\alpha+1)+s+s^{2^{k-2}}(\alpha+1)^{1-2^{k-2}}+s^{2^{k-2}+1}R(s)\\
&=& s^{2^{k-2}}(\alpha^{1-2^{k-2}}+(\alpha+1)^{1-2^{k-2}})+s^{2^{k-2}+1}R(s)
\end{eqnarray*}
which is a contradiction.

Suppose next that
$s=t=2^{2k-1}-2^{k-1}-1$ in which case
the degree $s$ equation is
\[
P_s Q_0+P_{0}Q_s=a_{s+3} \phi_{s+3}.
\]

If $Q_0=0$, then
$$\phi(x,y,z)=\sum_{j=3}^{d} a_{j}\phi_j (x,y,z)= (P_s+P_0) Q_t$$
which implies that
$$\phi(x,y,z)=a_{d}\phi_d(x,y,z)+a_{2^{2k-1}-2^{k-1}+2} \phi_{2^{2k-1}-2^{k-1}+2} (x,y,z)= P_sQ_t+P_0 Q_t$$
and  $P_0\ne0$, since $g\ne0$.
So one has
$\phi_{2^{2k-1}-2^{k-1}+2} $ divides $\phi_d(x,y,z)$
which is impossible by (\ref{decn}).

We may assume then that $P_0=Q_0$.
Then we have
\begin{equation}\label{phifactors}
\phi(x,y,z)=(P_s+P_0)(Q_s+Q_0)=P_sQ_s+P_0(P_s+Q_s)+P_0^2.
\end{equation}
Note that this implies $a_j=0$ for all $j$ except $j=3$ and $j=s+3$.  This means
\[
f(x)=x^d+a_{s+3}x^{s+3}+a_3x^3.
\]
So if $f(x)$ does not have this form, this shows that $\phi$ is absolutely irreducible.

If on the contrary $\phi$ splits as $(P_s+P_0)(Q_s+Q_0)$, the factors $P_s+P_0$ and $Q_s+Q_0$ are irreducible, as can be  shown by using the same argument.

Assume from now on that $f(x)=x^d+a_{s+3}x^{s+3}+a_3x^3$ and
that (\ref{phifactors}) holds.
Then $a_3=P_0^2$,
so clearly $P_0=\sqrt{a_3}$ is defined over $\mathbb{F}_q$.
We claim that $P_s$ and  $Q_s$  are actually defined over $\mathbb{F}_2$.

We know from (\ref{Goldfactors}) that $P_sQ_s$ is defined over $\mathbb{F}_2$.

Also $P_0(P_s+Q_s)=a_{s+3} \phi_{s+3}$, 
so $P_s+Q_s=(a_{s+3}/\sqrt{a_3}) \phi_{s+3}$.
On the one hand, $P_s+Q_s$ is defined over $\mathbb{F}_{2^k}$ by Theorem \ref{Kasami}. 
On the other hand, 
since $\phi_{s+3}$ is defined over $\mathbb{F}_2$ 
we may say that $P_s+Q_s$ is defined over $\mathbb{F}_q$.
Because $(k,n)=1$ we may conclude that
$P_s+Q_s$ is defined over $\mathbb{F}_2$.
Note that the leading coefficient of $P_s+Q_s$ is 1, so $a_{s+3}^2=a_3$.
Whence if this condition is not true, then $\phi$ is absolutely irreducible.

Let $\sigma$ denote the Galois automorphism $x\mapsto x^2$.
Then $P_sQ_s=\sigma(P_sQ_s)=\sigma(P_s)\sigma(Q_s)$,
and $P_s+Q_s=\sigma(P_s+Q_s)=\sigma(P_s)+\sigma(Q_s)$.
This means $\sigma$ either fixes both $P_s$ and $Q_s$, in which case we are done,
or else $\sigma$ interchanges them.
In the latter case, $\sigma^2$ fixes both $P_s$ and $Q_s$, so they are defined over
$\mathbb{F}_4$.
Because they are certainly defined over $\mathbb{F}_{2^k}$ by Theorem \ref{Kasami},
and $k$ is odd, they are defined over 
$\mathbb{F}_{2^k} \cap \mathbb{F}_{4} =\mathbb{F}_{2}$. 

Finally, we have now shown that $ X $ either is irreducible, or splits into two absolutely irreducible
factors defined over $\mathbb{F}_q$.
\x

\noindent
{\bf Remark}:
For $k=3$, the polynomial $\phi$  corresponding to
$f(x)=x^{57}+ax^{30}+a^2x^3$ where $a\in \mathbb{F}_q $ is irreducible.
Indeed if it were not, we would have $P_{27}$ and $Q_{27}$ defined over $\f2$, so by Theorem \ref{Kasami} we would have 
$P_{27}=p_\beta(x,y,z)p_{\beta^2}(x,y,z)p_{\beta^4}(x,y,z)$ and $Q_{27}=p_{\beta^3}(x,y,z)p_{\beta^5}(x,y,z)p_{\beta^6}(x,y,z)$ for some $\beta\in\f8-\f2$.
So, up to inversion, we would check that
$P_{27}(x,0,1)=(1+x+x^3)^9$ and $Q_{27}(x,0,1)=(1+x^2+x^3)^9$, hence
$P_{27}(x,0,1)+Q_{27}(x,0,1)=(1+x+x^3)^9+(1+x^2+x^3)^9$, and one can check that this is not equal to $\phi_{30}(x,0,1)$ as it should be.

\noindent
Institut de Math\'ematiques de Luminy, CNRS, Universit\'e de la M\'editerran\'ee, Marseille\\
e-mail: {\tt rodier@iml.univ-mrs.fr}

\end{document}